\documentclass[usenatbib]{mnras}

\usepackage{times}
\usepackage{graphicx}
\usepackage{amsmath}
\usepackage{amssymb}
\usepackage{color}
\usepackage{journal_names}
\usepackage{paralist}
\usepackage[inline]{enumitem}
\newcommand{\kms}{\mbox{\,km~s$^{-1}$}}
\title[The Fornax Cluster VLT Spectroscopic Survey]{The Fornax Cluster VLT 
Spectroscopic Survey. I -- VIMOS spectroscopy of compact stellar systems in the Fornax core region}

\author[Pota et al.]{\noindent
V. Pota$^{1}$\thanks{E-mail: vincenzo.pota@gmail.com}, 
N. R. Napolitano$^{1}$, 
M. Hilker$^{2}$, 
M. Spavone$^{1}$, 
C. Schulz$^{2}$,
Michele Cantiello$^{3}$,
\and 
C. Tortora$^{4}$,
E. Iodice$^{1}$, 
M. Paolillo$^{5,13}$,
R. D'Abrusco$^{6}$,
M. Capaccioli$^{5}$, 
T. Puzia$^{10}$ ,
\and 
R. F. Peletier$^{4}$,
A. J. Romanowsky$^{7}$,
G. van de Ven$^{2}$,
C. Spiniello$^{1}$,
M. Norris$^{9}$,
T. Lisker$^{8}$,
\and 
R. Munoz$^{10}$,
P. Schipani$^{1}$,
P. Eigenthaler$^{10,11}$,
M. A. Taylor$^{12}$,
\and
R. S\'anchez-Janssen$^{14}$,
Y. Ordenes-Brice\~no$^{8,10}$
\\~\\
$^1$ INAF -- Osservatorio Astronomico di Capodimonte, Salita Moiariello, 16, 80131 - Napoli, Italy\\
$^2$ European Southern Observatory, Karl-Schwarzschild-Stra{\ss}e 2, 85748 Garching bei Munchen, Germany\\
$^3$ INAF -- Astronomical Observatory of Abruzzo, Via Maggini, 64100, Teramo, Italy\\
$^4$ Kapteyn Astronomical Institute, University of Groningen, PO Box 800, 9700 AV Groningen, The Netherlands\\
$^5$ Dipartimento di Fisica Ettore Pancini, Universit\`a di Napoli Federico II,via Cintia, 80126 - Napoli Italy\\
$^6$ Harvard-Smithsonian Center for Astrophysics, 60 Garden St., Cambridge (MA), 02138, US\\
$^7$ University of California at Santa Cruz, Astronomy and Astrophysics Department, US\\
$^8$ Astronomisches Rechen-Institut Zentrum f{\"u}r Astronomie M{\"o}nchhofstra{\ss}e 12-14 69120 Heidelberg, Germany \\
$^9$ Jeremiah Horrocks Institute, University of Central Lancashire, Preston, PR1 2HE, United Kingdom\\
$^{10}$ Pontificia Universidad Catolica de Chile, Av. Vicuna Mackenna 4860, Santiago, Chile \\
$^{11}$ CASSACA, Camino El Observatorio 1515, Las Condes, Santiago, Chile\\
$^{12}$ Gemini Observatory, Northern Operations Center, 670 North A'ohoku Place, Hilo, HI 96720, USA\\
$^{13}$ INFN, Sezione di Napoli, via Cintia, 80126, Napoli, Italy\\
$^{14}$ UK Astronomy Technology Centre, Royal Observatory Edinburgh, Blackford Hill, Edinburgh, EH9 3HJ, U
}

\begin{document}

\label{firstpage}

\maketitle
\begin{abstract}
We present the results of a wide spectroscopic survey aimed at detecting 
extragalactic globular clusters (GCs) in the core of the Fornax cluster. 
About 4500 low resolution spectra (from 4800 
to 10000 \AA) were observed in 25 VLT/VIMOS masks covering the central 1 deg$^2$ 
around the dominant galaxy NGC~1399 corresponding to $\sim$175 kpc galactocentric radius. 
We describe the methodology used for data reduction and data analysis. We found a total of 
387 unique physical objects (372 GCs and 15 ultra compact dwarfs) in the field 
covered by our observations. Most of these objects are associated with NGC 1399, with only 
10\% likely belonging to other giant galaxies. 
The new VIMOS dataset is complementary to the many 
GC catalogues already present in the literature and it brings the total number 
of tracer particles around  NGC~1399 to more than 1130 objects. 
With this comprehensive radial velocity sample we have found that the velocity 
dispersion of the GC population (equally for red and blue GC populations) shows 
a relatively sharp increase from low velocity dispersion ($\sim 250-350 \kms$) 
to high velocity dispersion ($\sim 300-400 \kms$) at projected radius of $\approx10$ arcmin 
($\sim 60$ kpc) from the galaxy centre. 
This suggests that at a projected radius of $\approx60$~kpc both blue and red GC populations 
begin to be governed by the dominating Fornax cluster potential, rather than by the central 
NGC~1399 galaxy potential. This kinematic evidence corroborates similar results found using 
surface brightness analysis and planetary nebulae kinematics.
\end{abstract}


\section{Introduction}

Nearby galaxy clusters are ideal laboratories for studying the evolution of 
low and high mass galaxies as well as dense stellar systems (globular 
clusters, compact dwarf galaxies, planetary nebulae, etc.) in dense environments. 
The Virgo and Fornax clusters are the closest galaxy clusters, hence providing 
privileged targets where detailed observations of the galaxy and small stellar 
system content can be performed, and galaxy formation theories can be tested \citep[e.g.,][]{Strader11}. 

The Fornax cluster is the most massive galaxy overdensity after the Virgo cluster 
within 20 Mpc and it is an ideal target to study the effect of the environment 
on the structure and assembly of galaxies of any type, from the massive central 
giant early-type systems to the dwarf galaxies \citep[e.g.,][]{Ferguson89, munoz15, Iodice16, venhola17, Eigenthaler18}. Despite its regular appearance, 
it has been found that the assembly of Fornax is still ongoing. 
Although its core seems in an evolved phase \citep{Grillmair94,Jordan07} 
and most of the bright ($m_B < $  15 mag) cluster members are 
early-type galaxies \citep{Ferguson89}, the presence of stellar and GC tidal 
streams (e.g. \citealt{Iodice16}, \citealt{DAbrusco16}, Eisenhardt et al. 
2017) have revealed that there are still signs of active galaxy interactions in 
the region inside 200 kpc, which mirrors the large scale activity, including 
the accretion of the Fornax-A (NGC~1316) subgroup (dominated by the brightest 
cluster member, NGC~1316) into the cluster core along a 
cosmic web filament \citep{Drinkwater00,Scharf05}.

To kinematically map the complexity of the cluster core out to at least 200 kpc 
using discrete kinematical tracers (e.g. GCs, UCDs and PNe) and finally connect the 
large scale kinematics down to the scale of the dwarf galaxies, we have started 
a multi-instrument observational effort called the Fornax Cluster VLT 
Spectroscopic Survey (FVSS). As a part of this effort we have acquired integral 
field spectroscopy of dwarfs galaxies (see \citealt{mentz16}, Spiniello et al. 
in preparation), a counter dispersed imaging run with FORS2 to detect and 
measure radial velocities of planetary nebulae (Spiniello et al. 2018, 
submitted as FVSS II) and multi-object spectroscopy of GC and UCD candidates 
with VIMOS/VLT.

The latter program is the topic of this paper, where we present the radial 
velocity catalogue of GCs and UDGs in the core of Fornax, which is 
complementary to archival spectroscopic GC datasets on a more limited area: 
\citet{Schuberth} (which incorporates the catalogue of \citealt{Dirsch04}), 
\citet{Bergond07} and other catalogues discussed throughout the 
text. The novelty of the sample discussed in this paper is the uniform coverage 
of the central 1 deg$^2$ of the Fornax core which allows us to provide a 
spatially complete map of the kinematics of the galaxies in the field out to 
the regions where they meet the intracluster field (see e.g., \citealt{napolitano03}; 
\citealt{arnaboldi12}). Here, dynamical times are longer than the central 
galaxy regions and we expect to find the kinematical signature of the 
substructures already seen in the FDS photometry \citep{Iodice16} or other 
gravitational interactions such as shells and tidal tails in the kinematics of 
the galaxy outskirts or intracluster regions 
\citep{napolitano02,napolitano03, murante07, bullock05, rudick06, duc11, Longobardi15}. 
FVSS will be the foundation of a multifaceted project to constrain the baryonic and dark 
mass distribution in the core of the Fornax cluster with high precision, thus 
shedding light on the assembly history of massive galaxies in one of the 
nearest dense environments.

In this first paper of the series, we present the radial velocity catalogue of GCs. 
The assumed distance is 19.95 Mpc \citep{Tonry01}. Hence, $1$ arcmin $\sim 5.8$ Kpc.

The paper is structured as follows: Section \ref{sec:data} describes the data 
acquisition and reduction. The redshift estimation is discussed in Section 
\ref{sec:zest}, results are presented in Section \ref{sec:analysis} and 
discussed in Section \ref{sec:discussion}. We summarise the results of the paper and 
conclude in Section \ref{sec:conclusions}. 

\section{Data}
\label{sec:data}

This Section discusses all the steps undertaken to compile the spectroscopic 
catalogue discussed in this paper. The photometric surveys that constitute the input 
of our spectroscopic work are briefly described in \S\ref{sec:phot}. The 
workflow consists of selecting a sample of GC candidates based on photometric 
selection criteria (\S 
\ref{sec:selectionGC}) and then using the derived sample to design multi-object 
masks (\S \ref{sec:VIMOSpointings}) for follow-up 
observations (\S \ref{sec:observations}).

\begin{figure*}
\centering
\includegraphics[scale=0.7]{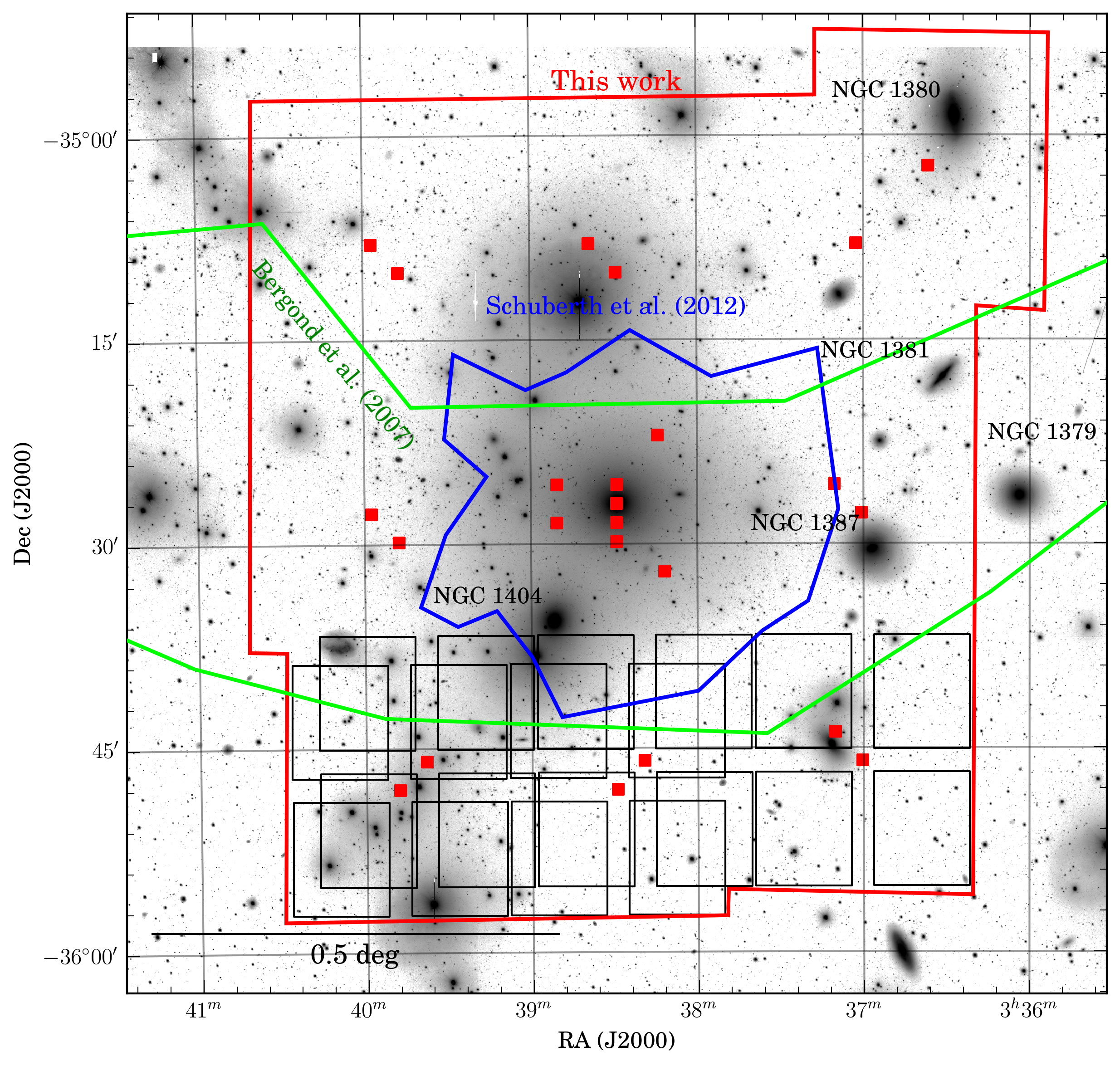} 
\caption{Layout of the observations. The background image is a mosaic of 
several VST/OmegaCAM pointings in the $g$-
band. The central galaxy is NGC~1399. Additional Fornax galaxies are also 
labeled. The red boxes mark the centers of all 
25 VIMOS pointings. The footprints of some VIMOS masks are shown on the bottom 
for illustrative purposes. Note how 
the VIMOS masks are dithered to maximize spatial coverage. The region 
covered by all 25 VIMOS is outlined in red, 
whereas the regions covered by literature studies are shown in green 
\citep{Bergond07} and blue \citep{Schuberth}.}
\label{fig:fov}
\end{figure*}

\subsection{Photometry}
\label{sec:phot}

\subsubsection{The Fornax Deep Survey}
\label{sec:FDS}

The Fornax Deep Survey\footnote{ESO programme ID (094.B-0687)} (FDS) is a joint 
project based on Guaranteed Time Observation surveys, FOCUS (P.I. R. Peletier) 
and VEGAS (P.I. E. Iodice, \citealt{capaccioli15}), aimed at studying the 
formation and assembly of the Fornax Cluster out to its virial radius with a 
variety of observations. These include
deep photometry, supplemented by cross-match with other wavelength observations 
such as  X-rays \citep[][e.g.,]{Paolillo02} and multi-object and integral-field 
spectroscopy.
The cornerstone of FDS is deep multiband ($u$, $g$, $r$ and $i$) imaging data 
from the OmegaCam (Kuijken 2011) camera at the VLT Survey Telescope (VST, 
\citealt{schipani12}), covering an area of $\sim30$ deg$^2$ of the cluster out 
to the virial radius. The area includes the Fornax-A subgroup, hence providing 
a full coverage of the cluster structure under assembly.  The depth of the 
imaging dataset reaches $\sim 30$ mag/arcsec$^2$ in the $g$-band \citep{Iodice16, Iodice17}, 
among the deepest acquired on the cluster so far \citep{venhola17}. 
VISTA/VIRCAM observations in $J$ and $K$ (with limiting magnitudes $K_s  = 23.4 $ $AB$ 
mag, and $J = 23.4$ $AB$ mag, respectively) are also available (see e.g. \citealt{munoz15}), to 
complement the optical data and optimise galaxy and globular cluster membership 
and the selection of follow-up targets (e.g. globular clusters, GCs hereafter, 
and/or ultra compact dwarf galaxies, UCDs hereafter). 

The multi-band deep images have been used to study the light distribution and 
colours of cluster galaxies out to 8-10 effective radii and beyond to 
characterise the faint galaxy haloes. These studies revealed the presence of 
ultra-faint stellar structures in the core of Fornax: fingerprints of past and 
ongoing interaction between galaxies falling into the deep potential well of 
the cluster \citep{Iodice16,venhola17}. The same conclusion is 
backed up by the complex distribution of globular clusters in the Fornax 
cluster \citep{DAbrusco16,cantiello17}.

\subsubsection{Next Generation Fornax Survey}
\label{sec:NGFS}
The Next Generation Fornax Survey (NGFS; \citealt{munoz15}, Puzia et al. 2018, in prep.) 
is an optical and near-infrared imaging survey of the Fornax
galaxy cluster virial sphere ($R_{\rm vir}\sim 1.4$~Mpc; \citealt{Drinkwater+01}). 
NGFS uses the Dark Energy Camera (DECam; Flaugher et al. 2015) mounted on the 
4-meter Blanco telescope at Cerro Tololo Interamerican Observatory (CTIO) in the 
optical wavelengths and the European Southern Observatory (ESO) 3.7-meter VISTA 
telescope with VIRCam \citep{sutherland15} for near-infrared observations. 
The current NGFS survey footprint covers $\sim50$ deg$^2$ with 19 DECam tiles of 2.65 
deg$^2$ each, and detects point-sources at $S/N\ =$ 5 down to $u'= 26.5$ mag , $g'= 26.1$ mag, 
$i'= 25.3$ mag , $J = 24.0$ mag and $K_s = 23.4$ mag. Here we make use of preliminary NGFS 
$K_s$-band point source catalogs.

The panchromatic NGFS data were recently used to detect low surface-brightness 
galaxies \citep{munoz15}, measure the morphological, scaling, and clustering 
properties of more than 600 dwarf galaxies (\citealt{Eigenthaler18}, îrdenes-Brice–o 
et al. 2018a, submitted), study their nucleation and stellar population properties 
(îrdenes-Brice–o et al. 2018b, submitted, Eigenthaler et al. 2018b, submitted), 
and analyze the globular cluster systems in Fornax ({\^I}rdenes-Brice–o et al. 2018c, submitted). 

\subsection{Selection of globular cluster candidates}
\label{sec:selectionGC}
The selection of GC candidates was based on VST/OmegaCAM photometry in the de-reddened $g$ 
and $i$ band from the FDS \citep{DAbrusco16,Iodice16} and preliminary 
VISTA/VIRCAM photometry in the $K_s$ band from the NGFS. 
In the combined 2-colour $g i K_s$ diagram, confirmed GCs from \citet{Schuberth}
are confined to a restricted colour-colour space. We defined a polygon around 
the radial velocity members in this colour space, which serves as first order 
selection criterion for our GC candidates. 

As an additional criterion we used the published wide-field Washington 
photometry from \citet{Dirsch04} and \citet{Bassino} to construct a $C-i$ vs. 
$i-K_s$ diagram. Since the Washington $C$-band is similar to a $u$-band filter 
and the $uiK_s$ plane is a very powerful tool to discriminate GCs from 
foreground and background objects \citep{Munoz14}, the $CiK_s$ plane is the 
cleanest selection criterion. Unfortunately, the spatial coverage of the 
Washington photometry is not complete in the central square degree due to chip 
gaps and the restricted field-of-view. Out of the 1065 unique spectroscopic 
targets with $CiK_s$ photometry, 809 (or $\sim$76\%) are within this selection 
polygon.

As shown in Sect. 2.2, the final 
mask design resulted in slit allocations on 4340 unique spectroscopic targets. 
4321 of those have $giK_s$ photometry, and 2643 (or $\sim61$\%) of them are 
contained in the selection polygon defined for potential GCs in the $giK_s$ space. 
We note that the missing u-band information in this colour space leads to a 
high fraction of contaminating foreground stars and background galaxies (see Sect. 3.1).

Here, we applied a magnitude restriction of 17.0$<i<$23.0 mag to our GC 
candidate sample in order to avoid severe contamination by foreground stars at bright 
magnitudes and too low signal-to-noise spectra at the faint magnitudes. 
For the very central regions of NGC~1399, which are dominated by the galaxy 
light, we found additional GC candidates taking advantage of the more accurate 
photometry and morphological classification derived from {\it Hubbe Space 
Telescope}/ACS data by \citet{Puzia14}.

Finally, all other allocated fibers, i.e. the remaining 1678 slits, besides a 
small portion used to duplicate targets, were split to observe known stars and 
background galaxies. These latter will be discussed in a separate paper. 

The FDS $gi$, the NGFS $K_s$, Washington $C$, as well as the central ACS  
photometric catalogs, were matched with SExtractor photometry of point and 
extended sources in all pre-images (ESO programme 094.B-0687), which were 
taken in $R$-band prior to the spectroscopic observations. The x and y 
coordinates of sources in the pre-images are needed to create VIMOS catalogs 
for the creation of mask files with the VMMPS (VIMOS Mask Preparation 
Software, \citealt{Bottini05}) from ESO.

\subsection{VIMOS pointings and mask design}
\label{sec:VIMOSpointings}

A total of 25 VIMOS masks were designed covering the central square degree 
around NGC~1399. The mask pointings were micro-dithered in such a way that the gaps between 
the four VIMOS CCDs (or quadrants) were covered by the adjacent mask. 

\begin{figure}
\centering
\includegraphics[width=\columnwidth]{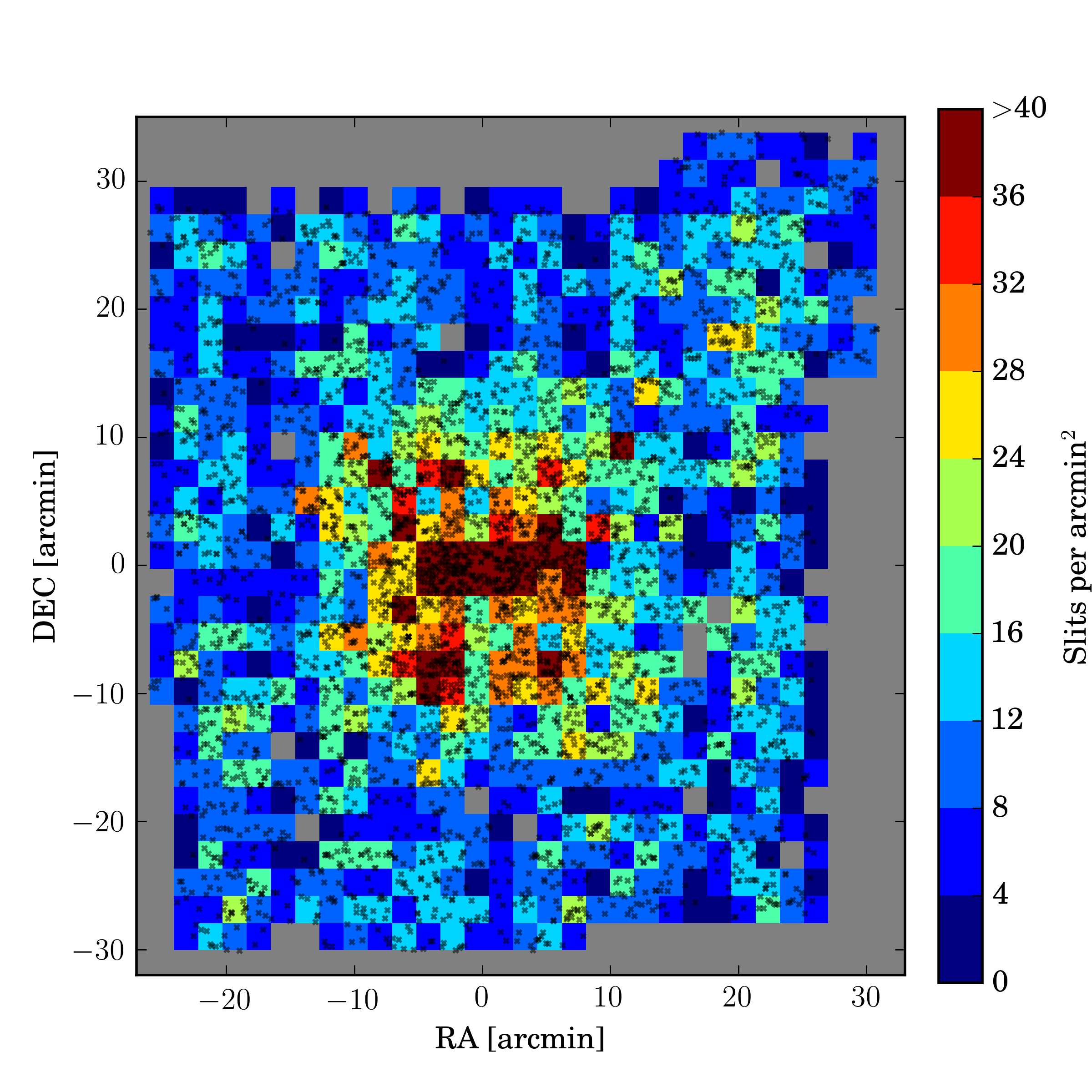} 
\caption{Two dimensional spatial density of VIMOS slits. Slits are shown as 
small black crosses, whereas the colours represent total counts in boxes of 
$2\times2$ arcmin$^2$. Axes express the distance in arcmin from NGC~1399. More 
slits were placed towards the center of the cluster, where the target density 
is higher. North is toward the top, East to the left.}
\label{fig:2dsplit}
\end{figure}

In Figure \ref{fig:fov} we show the spatial distribution of the VIMOS 
pointings, along with the footprints of archival observations in the same sky 
region from \citet{Schuberth} and \citet{Bergond07}. It is clear that the mask 
distribution is much more homogeneous with respect to previous studies, 
although our target sampling varies across the field. Besides the central giant 
elliptical NGC\,1399, the total covered area contains other giant galaxies, 
i.e., NGC~1404, NGC~1387 and NGC~1380.  
Figure \ref{fig:2dsplit} shows how the 2-dimensional density of allocated slits 
is considerably higher in the center with respect to the outskirts. Eight of 
the 25 pointings were dedicated to the central 25$\times$25 arcmin$^2$ to account 
for the higher density of GC candidates close to NGC~1399. 

The use of the MR grism allows a multiplexing of two in wavelength direction 
(parallel to declination), i.e. spectra are not overlapping if slits are placed 
close to the bottom and top of the chip areas in all quadrants. 
For a first automatic slit allocation we used VMMPS with a catalog of GC 
candidates from the $giK_s$ selection as input. Since the result of the 
automatic slit allocation was not satisfactory, we optimized the slit 
allocation of targets manually, for example by de-centering some targets in the 
slits while still keeping enough sky, or by allowing some overlap in wavelength 
range for multiplexed spectra, and by giving preference to targets that also 
fulfill the $CiK_s$ selection criterion. The remaining free area on the chips 
was then filled with slits centered on GC candidates from the $(g-i)$ colour 
selection or on pre-selected background galaxies to allow ancillary science on 
photometric redshift confirmations.

In this way, we defined 4574 slits in total for the 25 pointings ranging from 
157 to 202 slits per pointing (or 36 to 60 slits per quadrant). Several slits 
were positioned in such a way to cover more than one target if they had the 
same declination and a small distance in right ascension (of the order of 
10 arcsec or less). Since the 
fields are overlapping, about 300 targets have been observed twice and four 
targets even three times through different masks. This allows us to better 
understand uncertainties in the radial velocity measurements and correct 
for them between different masks. Discounting all duplications, we ended up 
with about 4340 unique targets for which spectra were taken. 


\subsection{Observations}
\label{sec:observations}

Spectroscopic observations were carried out with the VIsible MultiObject 
Spectrograph (VIMOS, \citealt{LeFevre}), mounted on the VLT-UT3 Melipal 
telescope at the ESO Paranal observatory in Chile and used in multi-object 
mode. Data were acquired in Period 94 (Program ID: 094.B-0687, 
PI: M. Capaccioli), from October 2014 to January 2015. 
The VIMOS spectrograph was equipped with a filter CG475, which cuts off 
wavelengths bluer than 4750 \AA\ , and a MR grating with a spectral resolution 
$R = 580$ (or 12.0 \AA\ FWHM) 
and a dispersion of $2.5$ \AA\/~pix. All slits had a width of 1 arcsec. The 
pixel scale in the spatial direction was 0.205 arcsec/pix. This setup allows us 
to explore the spectral window ranging from $4800$ to $10000$~\AA. Sky 
transparency was set to clear.

VIMOS is equipped with four CCDs, arranged in four quadrants with chip gaps of 
$\sim$2 arcmin in vertical and horizontal direction (see examples of the VIMOS 
detector footprints in Figure \ref{fig:fov}). Overall, one VIMOS mask set 
covers a field of 4 $\times$ (7 $\times$ 8 arcmin$^2$). All masks were observed 
for 1.5 hours in total, divided into three dithered exposures of 30 minutes. 
The seeing ranged from 0.66 arcsec to 1.15 arcsec, with a median of 0.85 arcsec. 
The median airmass was 1.061. 

\subsection{Data reduction}

The reduction of the VIMOS data was performed using the Reflex 
environment \citep{Freudling13} of the ESO VIMOS pipeline. 
For each VIMOS pointing, the data consist of the scientific images, flat fields and
arc lamp calibration images taken immediately after the science 
exposures. Each science exposure has a distribution of slits containing the 
science spectra of the targets. Indeed every slit is also contaminated by the 
emission spectrum of the earth's atmosphere. Since the wavelengths of the 
sky emission lines are precisely known they can be used for the absolute 
wavelength calibration of the spectra.

In order to obtain the final wavelength calibrated science spectra, we 
proceeded as follows. First, a wavelength calibration 
is performed using the provided arc lamp spectrum. Second, we calculated the residual 
shift of the sky emission lines with respect to their rest frame wavelength. 
For this purpose we provided our own sky line catalog, in which 
the wavelengths of prominent sky lines, in particular in the Ca triplet (CaT)
region, are listed. The pipeline determines this shift for each spectrum in 
each science exposure individually, which is essential because the wavelength 
shifts are not the same for the three exposures: the shift is largest for the 
first exposure and smallest for the third exposure. This 
is caused by the 
change of instrument flexure between the first science exposure and the final 
arc lamp exposure. In other words, the third science exposure is taken 
closest in time, i.e. under almost the same instrument conditions to the arc 
lamp exposure, and thus its wavelength calibration is the most accurate. 
Finally, the pipeline stacks the three individual science exposures and 
extracts sky subtracted object spectra from the combined science frame.

Unfortunately, the pipeline does not apply the wavelength 
shifts to the individual spectra before the stacking, meaning that the stacked
spectra are not corrected for the residual shifts. This leads to an incorrect
absolute wavelength calibration and also to a slight line broadening in the 
stacked spectra. Even though we cannot correct for the line broadening with
the current ESO pipeline version, we found a workaround to correct for the
absolute wavelength calibration of the spectra, which is necessary for an 
accurate measurement of the radial velocities of the objects. To zero-th order, 
it can be assumed that the median wavelength shift in the stacked spectrum is 
the same as the one derived for the second exposure. Thus, we re-reduced the 
second science exposure of all available data to determine the median 
wavelength shift in each multiplex of each mask by taking the average shift 
of all spectra in that multiplex. Then, the wavelength of each stacked 
spectrum was shifted by an amount as determined by the multiplex and the mask
in which the spectrum is located. All further analyses were carried out 
on the spectra corrected in this way.

For each mask, the REFLEX pipeline outputs multi-extension fits files which 
include the calibrated 2D spectra, the calibrated 1D spectra (and respective 
errors) and a 2D model of the sky lines. To make the dataset more manageable, 
we implemented a {\it python/astropy} \citep{astropy} script which copies different data structures 
corresponding to each slit into a multi-extension fits file. Each slit is 
associated to a fits file which contains: the 1D scientific spectrum, the 1D 
error spectrum, the sky-subtracted, wavelength-corrected 2D spectrum and a 
thumbnail of the object of $5\times5$ arcsec$^2$ from $g$-band VST imaging. 

\section{Analysis}

\subsection{Redshift estimation}
\label{sec:zest}

A total of 6700 spectra were extracted by the REFLEX pipeline. The number of spectra is larger
than the number of slits because multiple spectra can be extracted from one slit. 
These spectra belong to objects genuinely associated with the Fornax cluster (i.e., GCs, 
UCDs or dwarf galaxies), as well as Galactic stars, background galaxies and 
sources with signal-to-noise (S$/$N) too low to be classified with certainty. 

Radial velocities were computed with {\it iraf/fxcor}, which performs 
cross-correlation between the Fourier-transformed scientific spectrum and a 
Fourier-transformed set of template spectra \citep{Tonry79}. {\it Fxcor} was 
preferred to full spectra fitting approaches (e.g. pPXF, 
\citealt{Cappellari04}) because the former does not require initial guesses on 
the radial velocity and it is more CPU efficient. 

The Indo-U.S. Library of Coud\'e Feed Stellar Spectra \citep{Valdes04} was used 
as library of template spectra. The library contains 1273 stellar spectra, 
observed with a dispersion of $0.44$ \AA /pix and a resolution of $\sim 1$ \AA 
. The spectra cover the 3460 -- 9464 \AA\ range, which overlaps adequately with the 
wavelength range of our VIMOS spectra. From the whole library, we selected a 
random subset of 40 stars with spectral types from \textit{F} to \textit{M} 
because this is the range expected for metal-poor to metal-rich GCs. 

The 40 stellar spectra were convolved with a Gaussian filter of standard 
deviation $\sigma = 12.0\mbox{\AA\ } / 2.355$, where 12.0 \AA\ is the FWHM 
resolution of our VIMOS spectra and 2.355 is the constant relating the FWHM with $\sigma$ . 

The scientific spectra were prepared as follows. First, we computed the median 
signal-to-noise per resolution element $S/N$, where the signal $S$ is measured in the range 
5000-6600 \AA\ and the noise $N$ is the noise returned by the REFLEX pipeline 
in the same wavelength range. Second, telluric bands in the range 6850 -- 7688 
\AA\ and some bright skyline regions skylines were replaced with the fit to the spectral 
continuum. The continuum was computed with iraf/continuum by averaging ten 
contiguous pixels and interpolating the result with a cubic spline coupled with 
a $3\sigma$ rejection algorithm.  

Naively, one may think that cross-correlating the full wavelength range of our 
VIMOS spectra will return more robust radial velocities because the number of 
atomic lines used for the convolution is larger. Instead, we found that using 
the whole wavelength range can increase uncertainties due to severe template 
mismatches, likely due to the low ($S/N = 12$) average $S/N$ of our spectra. 
Therefore, we chose to run {\it fxcor} on the region surrounding the Calcium 
Triplet (CaT) at 8498-8548-8662 \AA , because these three lines occupy a very 
narrow wavelength range. Therefore, we correlate the scientific spectra between 
8485-8750 \AA\ with the template spectra between 8450-8720 \AA . This means 
that we are able to measure the radial velocities of unresolved objects between 
$-500 \kms$ and $+3000\kms$, which include both Galactic stars and Fornax 
objects.   

During the {\it fxcor} run, and prior to the correlation with the template 
spectra, the Fourier-transformed scientific spectra were filtered with a ramp 
filter. The filter was set up to cut off noisy high frequencies as well as low 
frequencies which may result from poor continuum fitting. The continuum was 
fitted with a cubic spline function.
We ran {\it fxcor} in non-interactive mode, meaning that {\it fxcor} always 
returns the velocity corresponding to the highest cross-correlation with the 
template spectrum. The final radial velocity is the median of the single radial 
velocities obtained from the cross-correlation with the 40 template stars. 
To overcome extreme template mismatch (i.e. the case in which the fitted radial 
velocity varies considerably depending on the adopted template), we used a 
median-absolute-deviation algorithm (MAD) to flag radial velocities deviating 
more than 5$\sigma$ from the median of the radial velocities obtained from the 
40 template stars. Outliers, if any, are removed from the velocity set and the 
final radial velocity and error are computed from the remaining measurements. 
90 per cent of GCs and stars have between 0 to 5 outliers, suggesting that the 
effect of extreme template mismatch is small. 

The scatter in radial velocity due to normal template mismatch (i.e. due to 
intrinsic mixture of stellar populations in GCs) varies between 3 -- 10 \kms 
(after removing outliers) depending on the $S/N$ of the spectrum. The final 
error on the radial velocity is the median velocity error returned by {\it 
fxcor} summed in quadrature to the scatter due to normal template mismatch. 

\begin{figure}
\centering
\includegraphics[width=\columnwidth]{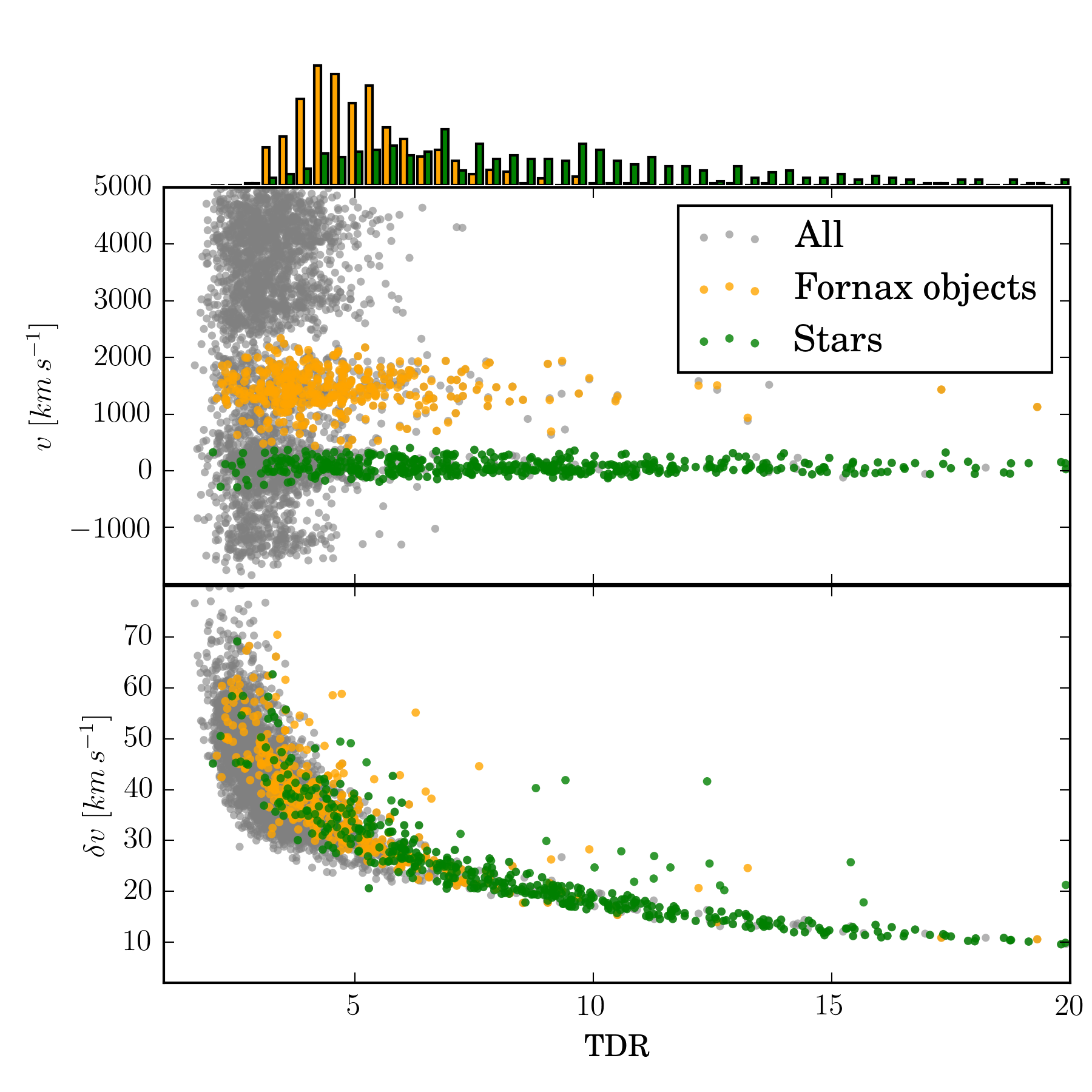} 
\caption{Results from {\it iraf/fxcor} for $\sim$ 4200 spectra. The x-axis 
represents the Tonry \& Davies R parameter (TDR) which is a proxy for the 
goodness of the cross-correlation between the scientific and the template 
spectra. Smaller TDR corresponds to larger velocity uncertainties $\delta v$ 
(bottom panel) and therefore to a poorer spectral fit. The measured radial 
velocity $v$ is shown on the y-axis of the top panel. The sequence of objects 
at $v\approx 0 \kms$ are Galactic stars, whereas the cluster at 
$v\approx1700\kms$ are sources in the Fornax cluster. The histogram on 
the top shows raw counts for stars and Fornax objects, respectively.}
\label{fig:tdr}
\end{figure}

\begin{figure}
\centering
\includegraphics[width=\columnwidth]{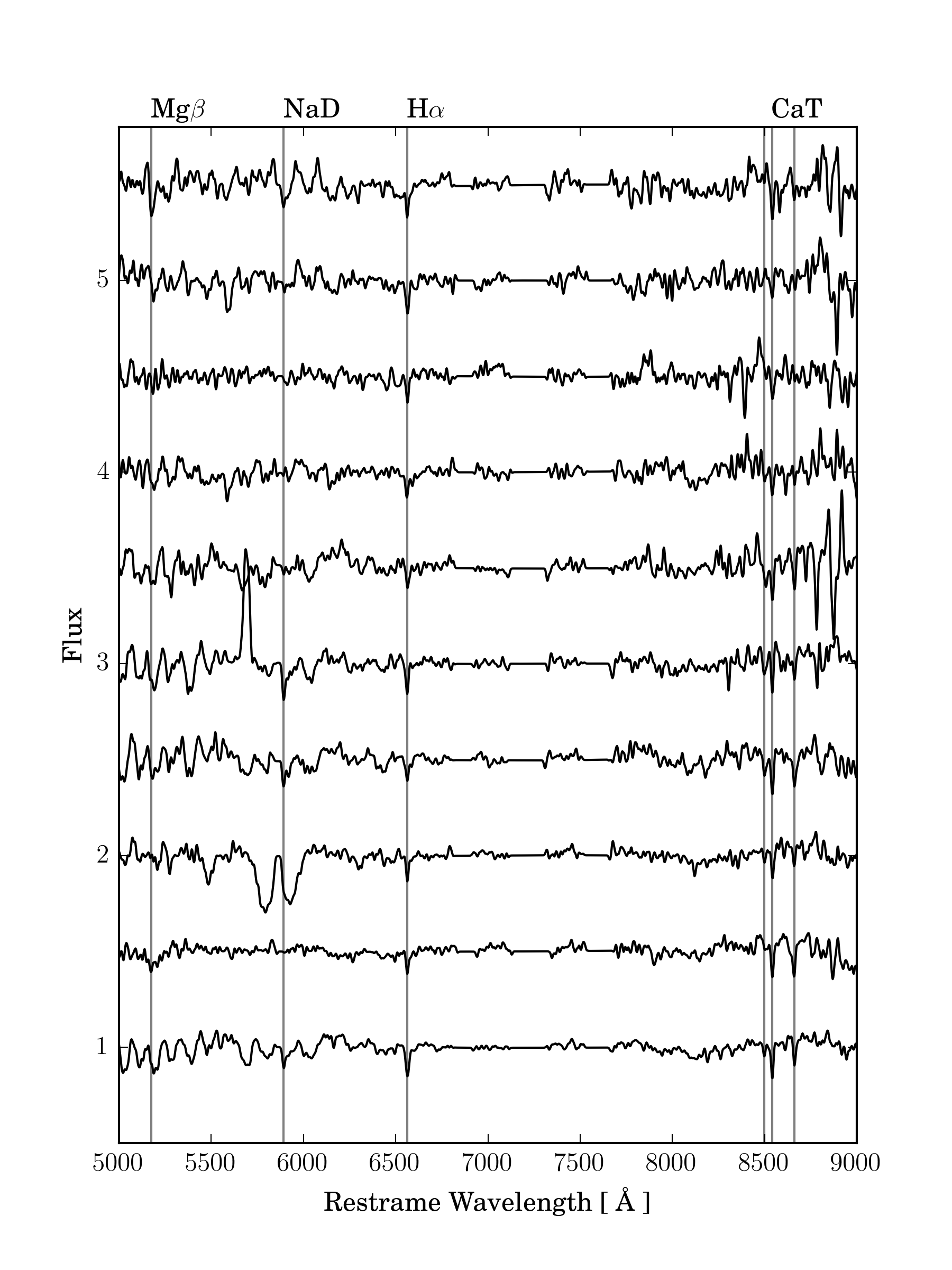} 
\caption{A sample of GC spectra suitably normalised. Shown are 
continuum-normalized redshift-corrected spectra for GCs with increasing 
$S/N$ per resolution element (from top to the bottom). The spectrum at the top has $S/N=15$, 
the one at the bottom $S/N=50$. The main atomic lines are labelled on the top. Features 
such as the strong absorption in the spectrum at Flux $= 2$ and the emission in 
the spectrum at Flux $=3$ are instrumental artifacts and not physical features 
of the GC. The flat horizontal lines between 7000-8000 \AA\ mark regions with 
masked skylines (see text).}
\label{fig:spectra}
\end{figure}

We used the \citet{Tonry79} R factor (TDR hereafter) to assess the goodness of 
the correlation between scientific spectra and template spectra. As expected, 
TDR is inversely proportional to the velocity uncertainties. This effect is 
shown in Figure \ref{fig:tdr}, where TDR is shown as a function of the measured 
radial velocity $v$ and velocity uncertainty $\delta v$. 
It is worth noting that {\it fxcor} was set to measure the Doppler shift of the 
CaT lines only for objects with radial velocities $-500 < v < 3000 \kms$. This 
means that only radial velocities measured to be $-450 < v < 2500 \kms$ are 
meaningful, whereas velocities outside this range are non-physical, and thus 
discarded as meaningless. Heliocentric velocity was computed with {\it iraf/rvcorrect} 
for each slit, and subtracted from all radial velocities. 

\subsection{Disentangling Fornax objects from foreground and background 
sources}

We are interested in separating objects physically bound to the Fornax cluster 
(GCs, UCDs and galaxies) from objects unrelated to the Fornax cluster (Galactic 
stars and foreground or background galaxies). A first distinction can be 
performed on the basis of the systemic velocity. We use the same velocity range 
adopted by \cite{Schuberth} for GCs and Galactic stars.
Using the top panel of Figure \ref{fig:tdr} as a reference, Fornax objects are 
those with $450 \le v <2500 \kms$ (850 datapoints), whereas Galactic stars are 
those with $-450<v<450 \kms$ (956 datapoints). The velocity determination of 
background galaxies and their distribution will be discussed in a separate 
paper.

Spectra of Fornax objects (GCs, UCDs) and Galactic star candidates were 
redshift-corrected and visually inspected. We checked for the correct positions of the 
following lines: CaT, H$\alpha (6563 \, \mbox{\AA})$, NaD$(5892 \, 
\mbox{\AA})$, the Mg$\beta triplet (\sim 5175 \, \mbox{\AA})$ and H$\beta (4861\, 
\mbox{\AA})$. The CaT and H$\alpha$ lines, if present, are always visible, 
whereas the Mg$\beta$ lines are hardly recognizable in spectra with $S/N \lesssim 
10$. The NaD is not always visible because this atomic line is shifted onto a 
problematic sky line at 5860--5890 \AA\ for spectra with $v \lesssim 1000\kms$ . 
The H$\beta$ line, at which the instrumental efficiency is less than $20$ per 
cent, can be discerned only for spectra with $S/N \gtrsim 20 $.

Figure \ref{fig:spectra} showcases some spectra with $S/N$ ranging from 50 
(bottom spectrum) to 15 (top spectrum). 
The GG475 filter of the first and third quadrant, coupled with the MR-grism, 
introduce an absorption feature between about $5600 < \lambda < 6300$ \AA\ 
(see the third spectrum from the bottom in Figure \ref{fig:spectra} for an 
example). This occurs because the GG475-filter is composed of 10-20\% in its 
weight by sodium oxide. Impure manufacturing likely caused this feature but it 
does not affect our analysis because the blue part of the spectra is not used 
for our velocity measurements. Strong emission features can occasionally appear 
in the blue part of the spectra (see the fifth spectrum from the bottom in 
Figure \ref{fig:spectra}), but they are attributed to zero order overlaps that 
are not corrected for by the Reflex reduction pipeline, rather than to physical 
phenomena associated with the light source. 

For spectra with very low $S/N$ the distinction between atomic lines and noise 
features becomes somewhat subjective. Therefore, all 850 Fornax object 
candidates were independently eyeballed by five members of the team. The 
inspection was performed on a dashboard including the 2D image of the source, 
the redshift corrected spectrum and attributes such as magnitude and S/N. We 
gave a vote of $1$ to spectra which were certainly Fornax objects and $0$ to 
non-Fornax object spectra. Votes from all members were summed-up for each spectrum. 
We found that 323 spectra received a vote of 5/5 
(meaning that these spectra were classified as Fornax objects by all members of 
the team), 348 spectra received a vote $\ge$ 4/5 and 420 spectra received a vote 
$\ge$ 3/5. In the followings, we will use the last set of 420 spectra as our 
final spectroscopic catalogue of the Fornax cluster. 

We repeated the above procedure to select a sample of spectroscopically 
confirmed Galactic stars with radial velocities in the range $-450<v<+450 
\kms$, finding a total of 492 Galactic stars.

Of the total $4574$ slits, we have allocated $\sim$2400 slits on GC/UCD candidates, 
$\sim 800$ slits on background galaxies (selected via a star galaxy separation 
based on their colors and stellarity index from SExtractor), and $\sim1000$ 
on stars. \citet{Schuberth} have already spectroscopically confirmed the brightest GCs
in the central regions of NGC~1399 and and we considered more valuable to focus on GC 
candidates not yet confirmed by previous studies, while still observing a few tens of known GCs to calibrate 
our velocity measurements against literature studies. The small recovery fraction we have 
obtained ($\sim 17\%$) is due to a number of factors: contamination
from stars and background galaxies in the fainter end of the GC 
luminosities, low S/N objects due to their poor spectral quality (see Figure \ref{fig:tdr}),
and observation at large galactocentric radii where the GC density is 
below 1 GC$/$ arcmin$^2$ (see Figure 15 in \citealt{Schuberth}). 

\section{Results}
\label{sec:analysis}
\subsection{Velocity self-consistency}

\begin{figure}
\centering
\includegraphics[width=\columnwidth]{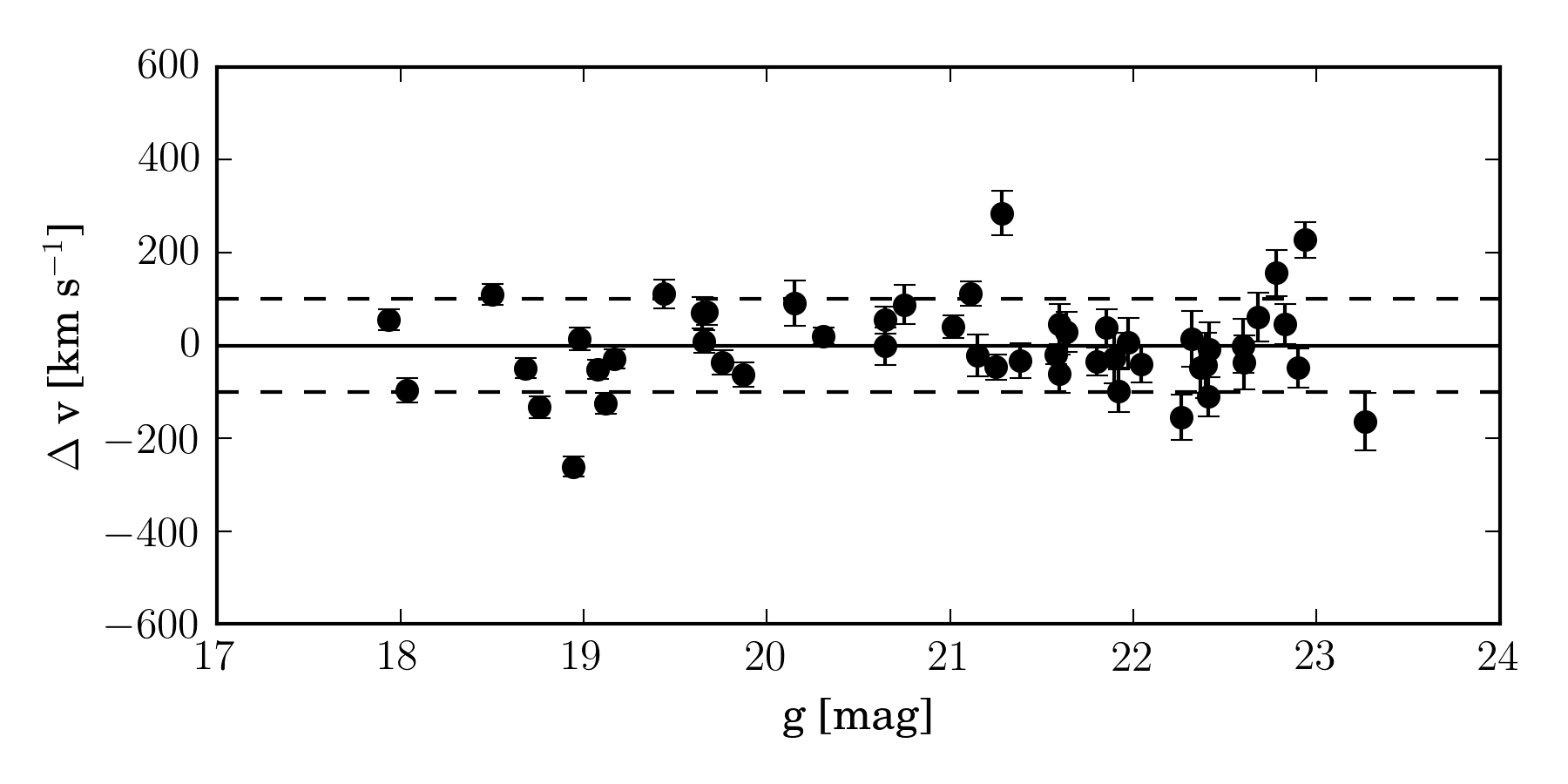} 
\caption{Internal velocity comparison. The figure shows the velocity difference 
$\Delta v$ between GCs with two independent velocity measurements. The solid 
and dashed line represent the reference $\Delta v = 0 \kms$ and 
$\Delta v = \pm 100 \kms$, respectively. }
\label{fig:internal}
\end{figure}

We measured radial velocities for 51 duplicated objects (including Galactic 
stars), observed across multiple VIMOS masks. The velocity difference $\Delta 
v$ between duplicated objects is shown in Figure \ref{fig:internal} as a 
function of $g$ magnitude. We detected no clear trend between $\Delta v$ and 
the magnitude, but a significant scatter is observed. One object has a velocity 
difference which scatters more than 
3$\sigma$ from the average. We attributed this disagreement to an issue with 
the wavelength calibration for this particular object. The root-mean-square of 
the velocity difference of GCs and stars combined is $91 \kms$, which becomes 
$79\kms$ after clipping the outlier. The former is our statistical error.

We weight-averaged the radial velocities of the duplicated GCs. This left us with a catalogue of 387 
Fornax objects and 464 Galactic stars.

\subsection{Comparison with literature data}\label{sec:compar}

The stellar systems surrounding NGC~1399 have been in the focus of many 
spectroscopic studies
\citep{Dirsch04, Schuberth, Bergond07, Firth07, Chilingarian11, Mieske04, 
Hilker07, Francis12, Drinkwater00}.

To find which of our VIMOS objects have a literature counterpart, we matched 
our catalogue of 387 Fornax objects with the NED (Nasa/Ipac Extragalactic 
Database\footnote{https://ned.ipac.caltech.edu/}) database using the 
{\it python/astropy} tools. 
We manually added the catalogue of \citet{Schuberth} to the literature 
database, because these GC measurements are not in NED. We required the 
literature objects to have a measured radial velocity and to be within 0.4 
arcsec from our VIMOS source. 

In most cases, the literature counterpart has been observed by more than one 
author. Table \ref{tab:authors} reports the number of objects per author for 
which we have a VIMOS radial velocity. 
In order to compare literature velocities with our VIMOS velocities, we 
averaged together the radial velocities of unique objects observed in multiple 
literature studies. 
Overall, we found 48 of our Fornax objects to have a literature counterpart. 

\begin{figure}
\centering
\includegraphics[width=\columnwidth]{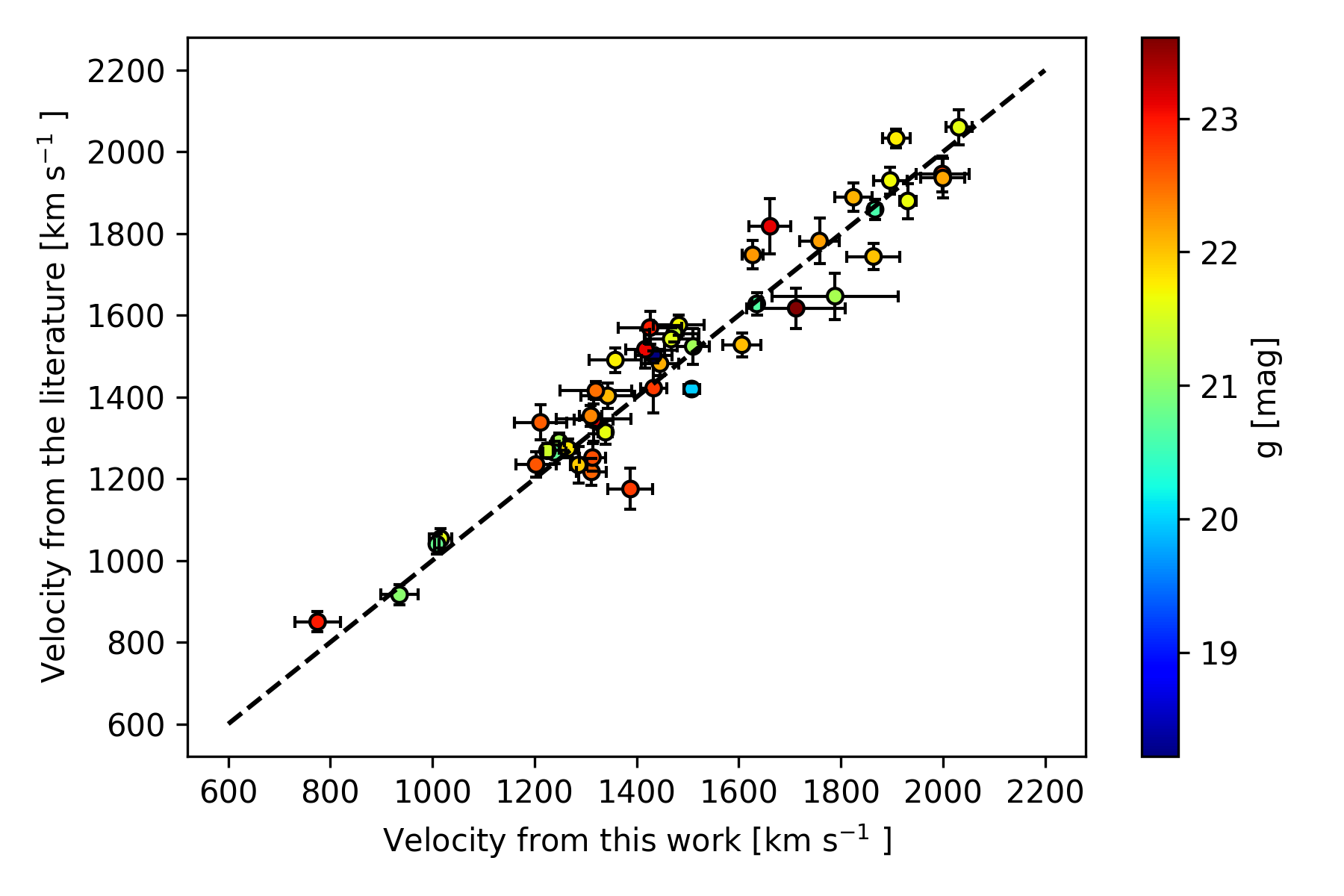} 
\caption{Comparison with literature studies. The radial velocity of literature 
objects is averaged among literature studies as discussed in the text. The 1-to-1 
line is shown as a dashed line. }
\label{fig:vel_comparison}
\end{figure}

Figure \ref{fig:vel_comparison} shows the comparison between our VIMOS 
objects and the unique literature objects obtained as explained above. No 
bias is detected. The root-mean-square of the velocity difference is $81 \kms$. 



\begin{table}
\centering
\label{mathmode}
\begin{tabular}{@{}l c}
\hline
Paper & Number of duplicated objects \\
\hline
\citet{Dirsch04}       &  29 \\
\citet{Schuberth}      &  26 \\
\citet{Bergond07}      &  14 \\
\citet{Firth07}        &  11 \\
\citet{Chilingarian11} &  2  \\
\citet{Mieske04}       &  1  \\
\citet{Hilker07}       &  1  \\
\citet{Francis12}      &  1  \\
\citet{Drinkwater00}   &  1  \\
\hline
\end{tabular}
\caption{Number of confirmed VIMOS objects also found in archival data grouped 
by author.}
\label{tab:authors} 
\end{table}

\subsection{Classification of Fornax objects}

We classified all Fornax objects based on their morphology: namely GCs, ultra 
compact dwarfs (UCDs) and background galaxies. The spatial resolution of 
VST/OmegaCam is insufficient to resolve the typical half-light radii of GCs 
with $r_h < 10$ pc \citep{Masters, Puzia14}. Although the size of UCDs with 
$r_h > 20$ pc can be resolved using our imaging data \citep{Cantiello15}, 
measuring UCD sizes is not the focus of this paper and will be deferred to a 
future work.

It is established that a sharp magnitude cut cannot perfectly separate UCDs 
from GCs. However, it has been shown \citep{Voggel16, 
Eigenthaler18} that the bulk of candidate UCDs with confirmed sizes in the Fornax cluster 
have $M_V \le -10$ ($\approx M_i \le -11$). 
Therefore, we use this magnitude cut as benchmark value to separate UCDs with 
$i \le 20.3$ from GCs with $i > 20.3$. This selection returns 15 UCD candidates 
and 372 likely GCs. 

The object with coordinates RA(J2000) $=$ 3:36:37.253 DEC(J2000) $=$ 35:23:09.20 is the nucleus of the 
nucleated dwarf galaxy FCC~171, whose radial velocity was first measured in 
\citet{Bergond07}. This object will be included in the GC sample in the following 
analysis.  

\section{Discussion}
\label{sec:discussion}
In this section we discuss some qualitative properties of the final dataset 
of bona-fide GCs and UCDs. We look at the spatial distribution and 
phase-space diagrams. A detailed kinematic and dynamic analysis 
(rotation curve, dark matter modelling, luminosity function analysis) will 
be the subject of forthcoming papers.

\begin{figure*}
\centering
\includegraphics[width=\columnwidth]{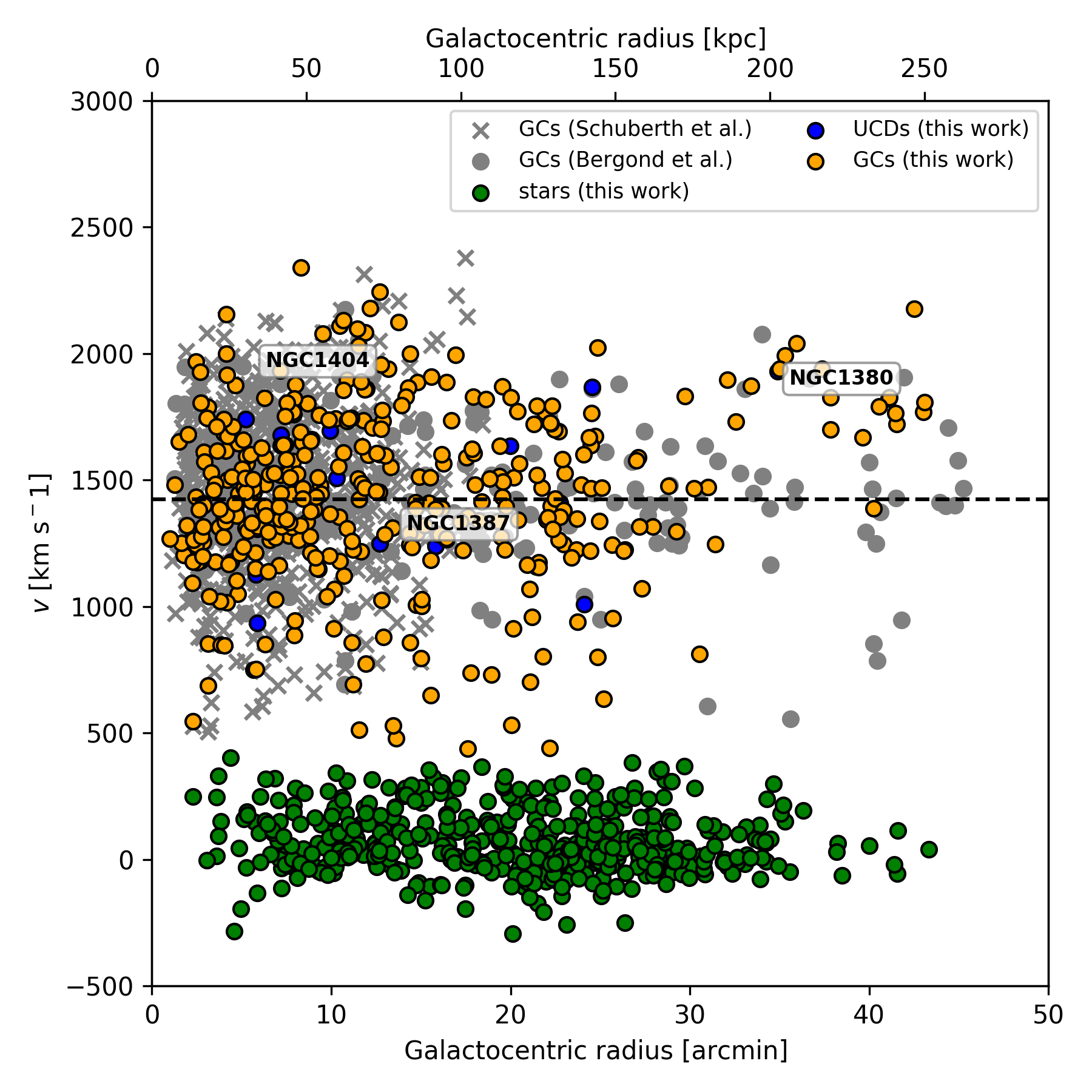} 
\includegraphics[width=\columnwidth]{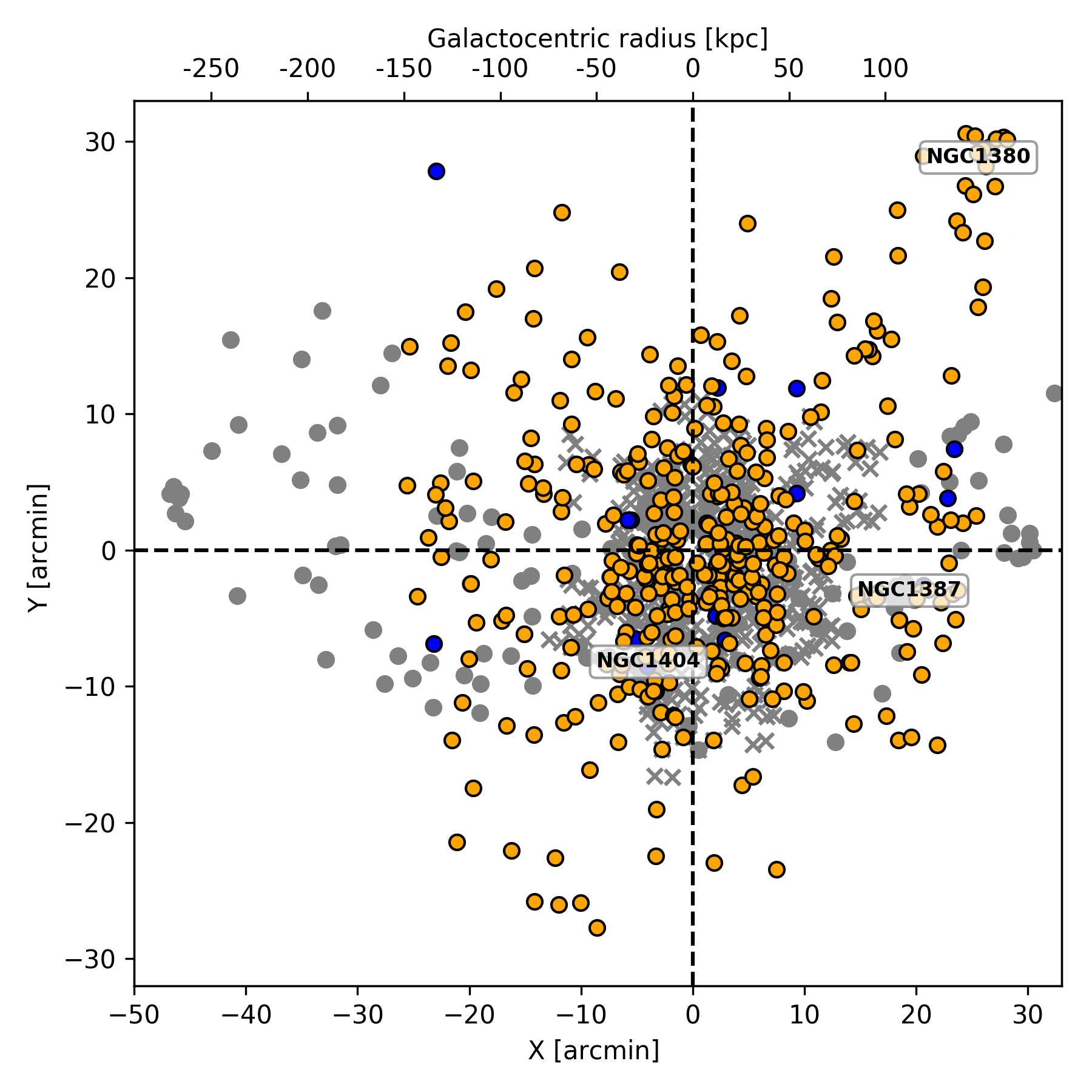} 
\caption{\textit{Left}: Phase-space velocity diagram. The figure shows the 
measured systemic velocity of the Fornax objects presented in this paper 
(orange points), GCs from \citet{Schuberth} (grey crosses), GCs from 
\citet{Bergond07} (grey points) and Galactic stars measured in this paper 
(green points). The systemic velocity of NGC~1399 is marked with a dashed line 
at $v = 1425$ km s$^{-1}$. The systemic velocity and galactocentric distance of 
giant Fornax galaxies (namely NGC~1404, NGC~1387, NGC~1380) are marked with 
the correspondent galaxy name.  The distance is expressed in 
arcminutes from the centre of NGC~1399. 
Physical distances in kpc are also shown on the top panel. 
\textit{Right}: Position diagram, with North to the top and East to the left. 
The plot shows the spatial distribution of the objects from the left panel 
(except for Galactic stars for clarity). Note how the GCs from our paper complement the GC 
catalogues of \citet{Schuberth} and \citet{Bergond07}. }
\label{fig:phase-space}
\end{figure*}

\subsection{Phase-space and spatial distribution}

The phase-space diagram and the spatial distribution are shown in Figure 
\ref{fig:phase-space}. In both cases, GCs, UCDs and Galactic stars are compared 
to the catalogues of \citet{Bergond07} and \citet{Schuberth}. These two 
catalogues are the largest and most homogeneous in the literature. 
Moreover, 
the catalogue of Schuberth is an extension of the catalogue of 
\citet{Dirsch04}. Together they include sources within 18 arcmin from NGC~1399. 
Bergond et al.'s catalogue was designed to target intra-cluster globular clusters and 
it is more radially extended (see also Figure \ref{fig:fov}). It covers a strip 
of about 1.5 degree in right ascension and half a degree in declination. The 
catalogues of Schuberth et al. and Bergond et al. combined provide a good representation of 
the current archival phase-space and spatial distribution of the GC system in 
the core of the Fornax cluster. 

The left panel in Figure \ref{fig:phase-space} shows that the velocity 
distribution of GCs/UCDs is well separated from that of Galactic stars, 
although a few GCs between $10<R<20$ arcmin 
might be misclassified as Galactic stars. 

The systemic velocity of GCs and UCDs is $v_{\rm GCs} = 1443 \pm 18 \kms$ and 
$v_{\rm UCDs} = 1413 \pm 91 \kms$, respectively. These are both consistent with 
the average systemic velocity from the literature $v_{\rm NED} = 1425 \pm 4 
\kms$. Our sample also includes interlopers from large galaxies surrounding 
NGC~1399, in particular NGC~1380, NGC~1404, NGC~1387. The positions of these 
three galaxies in the phase-space diagram are also labelled. 

It is not trivial to disentangle the GC population of NGC~1387 and NGC~1404 
because of their proximity in phase-space to NGC~1399. 
Here we adopt a conservative approach to extract the GC system of these 
galaxies, but we acknowledge that more refined methodologies can return more 
accurate results, as we will discuss in forthcoming papers.
Following \citet{Schuberth}, we use 3 arcmin from the galaxy 
centre as a benchmark radius to select the bulk of GCs associated with these 
galaxies. We also require that the radial velocities of GCs have to be within 
$v_{\rm sys} \pm 2\sigma$  the systemic velocity of the host galaxy, where 
$\sigma$ is the galaxy stellar velocity dispersion. Here we used $\sigma = 247 
\kms$  \citep{Vanderbeke11}, $v_{\rm sys} = 1947 \kms$ for NGC~1404 and $\sigma 
= 170 \kms$ \citep{Wegner03}, $v_{\rm sys} = 1302\kms $ for NGC~1387, 
respectively. 

After applying the above selection criteria, we found 17 and 8 
GCs associated with NGC~1404 and NGC~1387, respectively. 
The larger number of 
GCs associated with NGC~1404 is due to the galaxy being more massive than 
NGC~1387, but also to its proximity to NGC~1399 which increases the fraction of 
contaminants.
In the case of NGC~1380, with $\sigma = 190 \kms$  \citep{Vanderbeke11} and 
$v_{\rm sys} = 1877 \kms$, the identification of its GC system is easier given 
its isolated position. Using the criteria above, we found 7 GCs associated with 
this galaxy.

The right panel of Figure \ref{fig:phase-space} displays the spatial 
distribution of our GC and UCD catalogues. This diagram also shows how our 
catalogue is complementary to literature catalogues: it increases the object 
sampling in the central regions of the galaxy and it fills gaps without literature data in 
the outer regions. 

GC systems which are bound to other giant galaxies can be seen clustered around 
their centres. Figure \ref{fig:phase-space} suggests that our combined sample 
might also contain some GCs belonging to NGC~1379 and NGC~1381 (see Figure 
\ref{fig:fov}). 

\begin{figure*}
\centering
\includegraphics[width=\columnwidth]{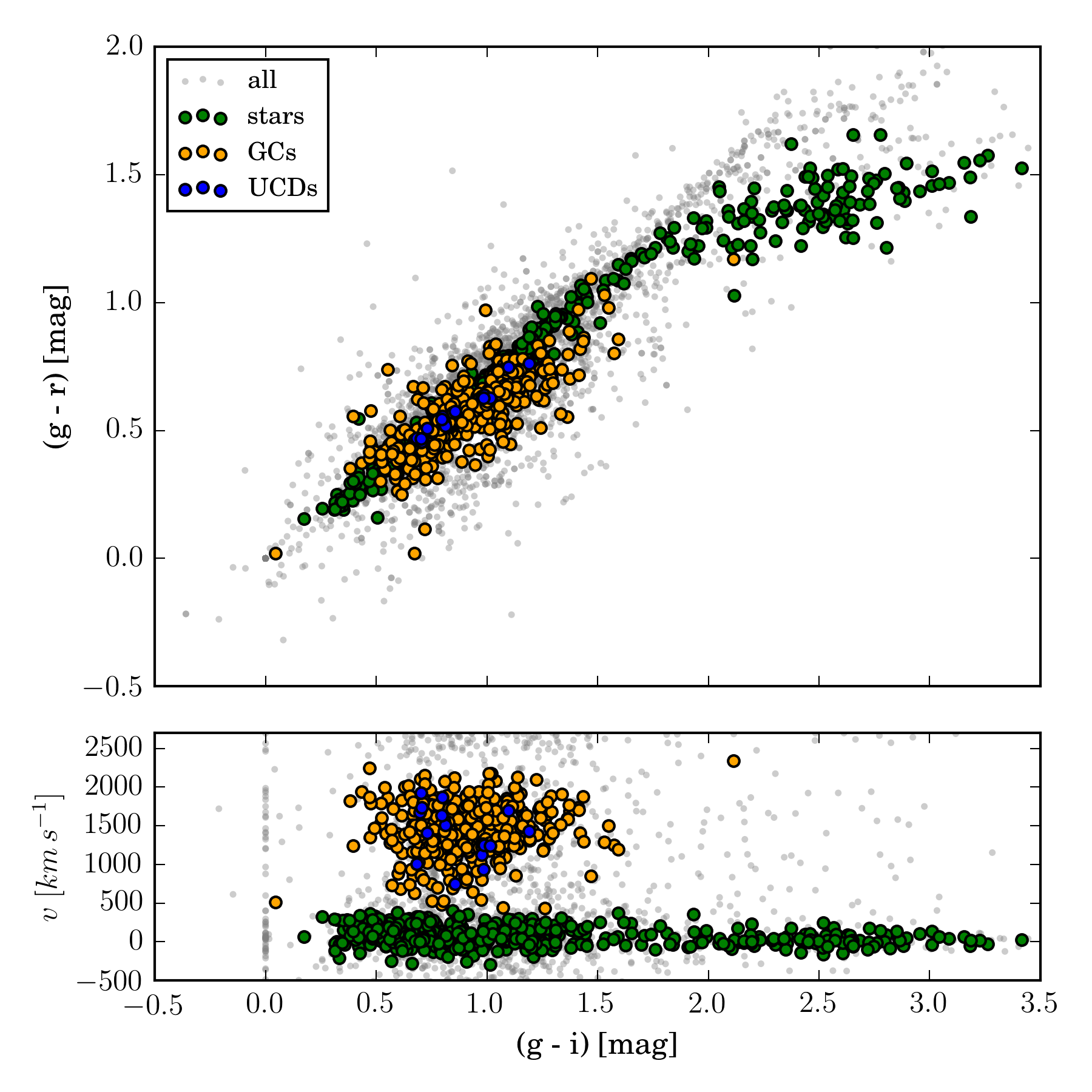} 
\includegraphics[width=\columnwidth]{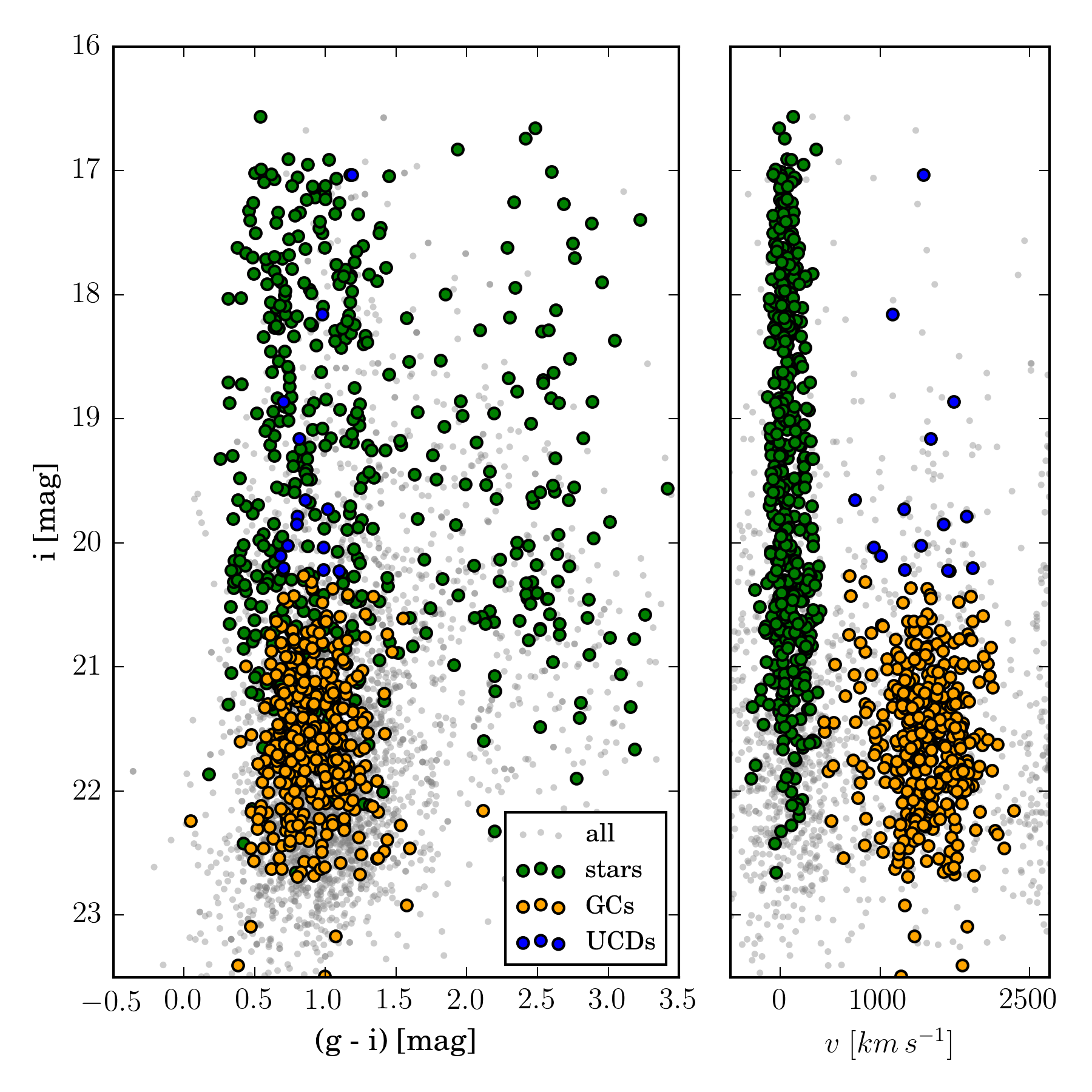} 
\caption{Colour--colour and colour--magnitude diagram. 
The figure shows the 
distribution of the labelled objects in a $(g-i)$ vs. $(g-r)$ space (left 
panel) and $(g-i)$ vs $i$ space (right panel). Each diagram includes also a 
radial velocity $v$ axis to show the separation of GCs/UCDs from stars and 
galaxies. The objects labelled as `all' represent all 4340 sources observed with VIMOS.
}
\label{fig:cc}
\end{figure*}

\subsection{Photometric diagrams}


Figure \ref{fig:cc} displays the distribution of GCs/UCDs in the photometric 
space. These are compared with the distribution of all sources targeted for 
spectroscopic follow-up, including Galactic stars and background galaxies. 
Although the $u$ filter is known to well discriminate GCs from contaminants, 
here we only consider $g$, $r$, $i$ magnitudes because all but one GCs/UCDs 
have a genuine measurement in these photometric bands, whereas only 50\% of our 
GCs/UCDs have a FDS $u$-band measurement. This is due to FDS $u$-band imaging being 
shallower than the other bands for detecting faint GCs \citep{DAbrusco16}.

GCs/UCDs and contaminants are not well separated using merely $g$, $r$, $i$ 
filters, but velocity information clearly separates GCs/UCDs from Galactic 
stars and background galaxies. This is also shown in Figure \ref{fig:cc} 
(left-bottom panel with $(g-i)$ vs $v$) which shows how the velocity 
dispersion of blue GCs with $(g-i) \le 0.85$ is higher than that of the red GCs, a 
property shared with GC systems in the most giant galaxies \citep[e.g.,][]{Pota13}. 
The rightmost panel, with $v$ vs $i$, shows that 
the velocity distribution of all GCs has an extended tail towards lower velocities (a 
property also visible in Figure \ref{fig:phase-space}). This effect was already 
noticed in \citet{Schuberth}. The convincing separation between GCs and stars 
also ensures that this property is unlikely due to miss-classified GCs. 

\begin{figure*}
\centering
\includegraphics[scale = 0.8]{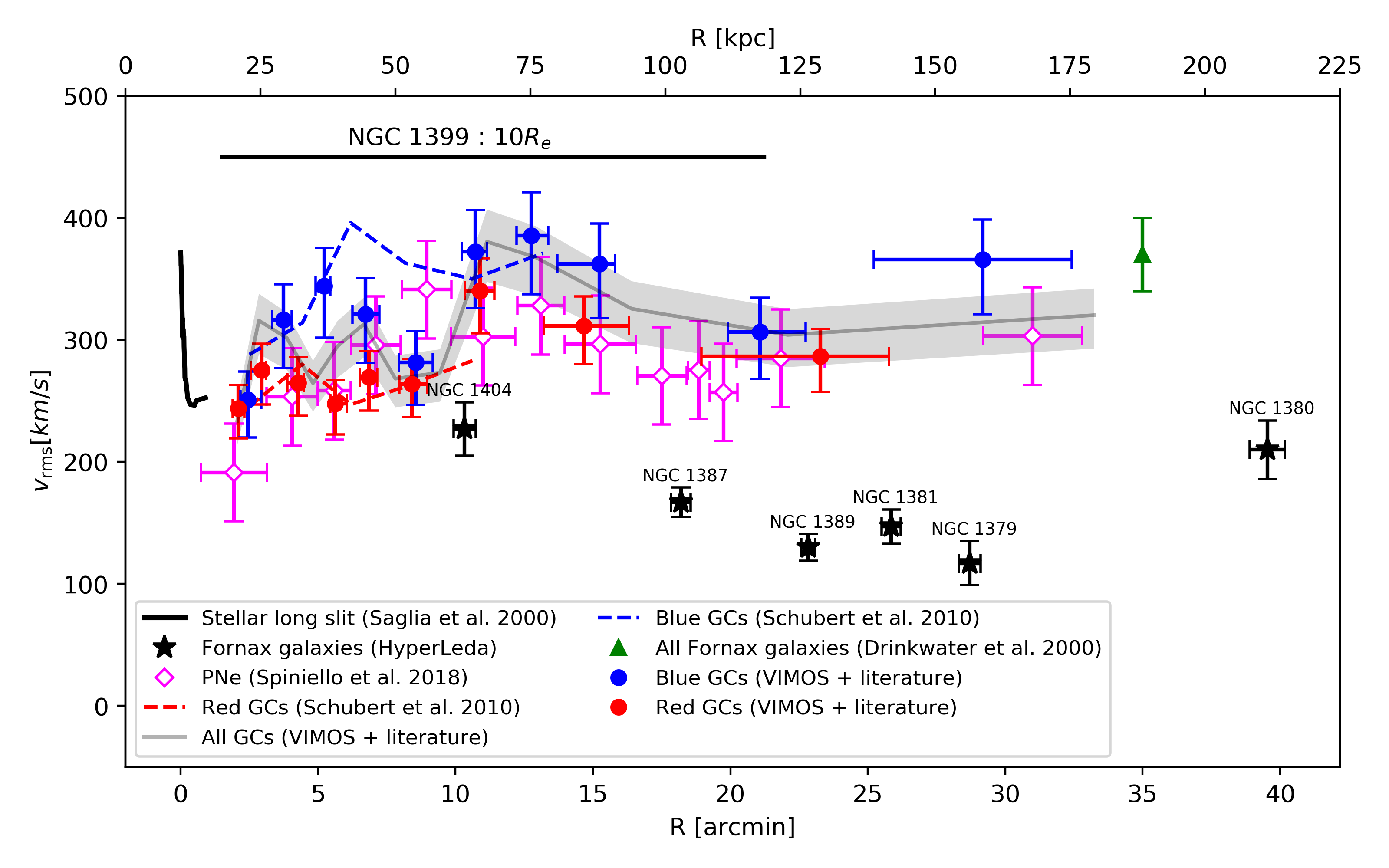} 
\caption{Root-mean-square velocity dispersion of NGC~1399 stellar system. The 
blue and red points represent the blue and red GCs/UCDs, respectively, whereas 
the full grey-line shows the full master catalogue of 1133 GCs/UCDs. Bins are 
of irregular size, reflected in the asymmetric radial error bars representing 
the 25-th and 75-th quantiles. The digitised $v_{\rm rms}$ profiles from 
\citet{Schuberth} are the blue and red dashed lines, without uncertainties for 
clarity. The central stellar velocity dispersion for the galaxies in the field 
are the black points, with the error bars on the x-axis representing one 
effective radius. The green triangle shows the innermost value of the velocity 
dispersion derived for Fornax galaxies by \citet{Drinkwater00}.}
\label{fig:vrms}
\end{figure*}

\subsection{Root-mean-square velocity profile}
The root-mean-square velocity $v_{\rm rms}$ quantifies the total kinetic energy 
of a stellar system. It accounts both for the ordered and for the random 
stellar motions. Here we compute the root-mean-square velocity as:
\begin{equation}
v_{\rm rms}^2 = \frac{1}{N}  \sum (v_i - v_{\rm sys})^2 - (\Delta v_i)^2 ,
\label{eq:sigma}
\end{equation}
where $v_i$ is the radial velocity of the i-th GC and $\Delta v_i$ is its 
uncertainty. $v_{\rm sys}=1425 \kms$ is the systemic velocity of NGC~1399. 
Uncertainties were derived with the formulae provided by \citet{Danese}. 
Statistical errors are discussed in Section \ref{sec:analysis}.

The $v_{\rm rms}$ has been calculated using a total of 1133 unresolved objects 
(including GCs and UCDs). Although we cannot exclude that GCs and UCDs are 
decoupled populations, 
the fraction of classified UCDs is too modest to affect the overall average 
kinematics of the combined populations. A detailed analysis of UCDs is beyond 
the purpose of this paper,
however, looking at Fig. \ref{fig:phase-space}, object classified as UCDs do 
not show any evident systematic difference with respect to the GC systems. 
The final catalogue is made of the 387 objects discussed in this paper 
combined with the catalogues of \citet{Bergond07} and \citet{Schuberth}. Our master catalogue
includes GCs from other giant galaxies in the field. If an 
object was found in more than one catalogue we computed its weight-average radial 
velocity. Also, GCs and UCDs from 
literature catalogues were given the VST photometry discussed in this work. 
Therefore, no photometric calibration was required.

Figure \ref{fig:vrms} displays the $v_{\rm rms}$ profile for: i) our combined master 
catalogue of 1133 GCs and UCDs, including literature objects; ii) the GC catalogue of 
\citet{Schuberth}, iii) the central stars of galaxies surrounding NGC~1399; iv) 
all Fornax galaxies from \citet{Drinkwater00}, v) PNs from FVSS-II (Spiniello et al., 2018, submitted).
The $v_{\rm rms}$ profile is computed in radial bins of irregular 
size. However, each bin contains roughly the same number of objects, namely: 
48, 66, 88 objects per bin for the blue, red and all GCs (including UCDs), respectively. 

The master catalogue contains GCs and UCDs predominantly from NGC~1399, which 
is the largest and most massive system in the field dominating the total 
potential in the core of the cluster. Here, objects from other galaxies, 
embedded in the extended exponential halo around the galaxy \citep{Iodice16}, 
in particular NGC~1404, NGC~1387, are likely to contaminate the 
kinematics of the central galaxy. However, looking at the $v_{\rm rms}$ 
profile of the total sample (shaded gray area) there is a steep rise in the 
profile around $R\sim10$ arcmin ($\sim 50$ Kpc) which marks a kinematical transition from a 
``colder'' kinematical region to a ``warmer'' one. 
The same behavior can be observed in the PNs datapoints 
(magenta triangles). As discussed in FVSS-II, the peak in $v_{\rm rms}$ 
of the point around $R\sim 8$ arcmin, could partially be due to NGC~1404. 
In fact, Planetaries bounded to this galaxy could possibly contaminate the sample, 
causing an increase of the $v_{\rm rms}$ in that particular radial bin.

Regarding the data form Schuberth et al., we use their full data dataset (corresponding to 
sample BI and RI in their Table A.1) because its properties are akin to those of our master catalogue.
The $v_{\rm rms}$ profile is in general agreement for the red GCs, but we
observe a disagreement for blue GCs in the region $5<R<10$ arcmin, where the data from 
Schuberth et al. are systematically higher than our datapoints. Such a discrepancy
might be due to the different colour threshold used for the blue/red separation: we
use $(g-i) = 0.85$ mag, whereas Schubert et al. use $(g-i) = 0.91$ mag (after converting
their Washington magnitudes to SDSS magnitudes). Also, we applied no magnitude cut before
calculating the $v_{\rm rms}$, whereas Schubert et al. only considered GCs brighter than $R>21.1$ mag.

In the outermost region, the $v_{\rm rms}$ profile flattens out to a value which is 
fairly consistent with the kinematics of the cluster galaxies (the green data 
point at $R\sim35$ arcmin from \citealt{Drinkwater00}). This suggests that the GC 
kinematics at radii $R>10$ arcmin is ruled by the cluster potential. 
This feature demonstrates that the photometric transition radius 
between the bright central galaxy and the outer exponential halo at $R\sim10$ 
arcmin from \citet{Iodice16} represents the radius where we kinematically observe the 
emergence of an intracluster population of GCs and PNe. 
At the light of similar conclusions found for PNs by Spiniello et al.(2018, submitted), 
there is  growing evidence that multiple independent lines of investigation 
are starting to support, for the first time, the existence of a photometrically 
and kinematically distinct intracluster component in the Fornax Cluster core.

When split in the red and blue subpopulations, using a colour cut at $(g - i) = 0.85$ mag 
the GC sample shows some interesting differences. At all radii the $v_{\rm rms}$ 
profile of the red GCs is systematically smaller than the one of the blue GCs, 
albeit the two profiles are broadly consistent with each other within one sigma. 
The $v_{\rm rms}$ of PNs is consistent with that of red GCs, as observed is 
most giant galaxies \citep[e.g.,][]{Pota13}.
In the innermost regions at $R < 10$ arcmin our results are consistent with  
those from \citet{Schuberth}. In particular the blue GCs turn out to 
be the more extended population in radius, nicely connecting to the value of 
the cluster galaxy datapoint. The difference in normalisation between the two 
populations may be driven by the different slopes of the two 
populations' density profiles as measured, e.g., by 
\citet[][their Eq. 10 and 11 and Table 5]{Schuberth}. 

Finally, we report in the same Figure the central velocity dispersions of 
the giant galaxies populating the same area occupied by the total GC catalogue (see Figure \ref{fig:fov}). 
For NGC~1399, we show long slit data from \citet{Saglia00} up 
to 1 arcmin galactocentric radius, where the stellar velocity dispersion is consistent with
the velocity dispersion of the innermost GCs.
At larger distances one can expect that neighbouring galaxies might
might contaminate the overall GC kinematics. Especially in the outer regions, 
i.e. $R>10$ arcmin, assuming these regions are dominated by the cluster 
potential, the presence of GCs bound to the galaxies might alter the true  
intracluster population. A detailed separation of the bound GC population from 
the true intra-cluster GCs (unbound from any galaxy in the core) will be addressed in a 
forthcoming paper. 

In conclusion of this section, we can summarise the main finding of this paper 
in Figure 9, by saying that the unprecedented extension of the GC kinematics, 
combined with the literature data, has demonstrated the kinematical signature 
of a population of intracluster GCs, made of both red GCs and blue GCs. 
The two populations show $v_{\rm rms}$ profiles consistent with the transition of GC 
dynamics driven by the potential of the central giant elliptical in Fornax (NGC~1399) 
in the innermost regions to GC dynamics governed by the cluster potential 
beyond $\sim$10 arcmin ($\sim$50 kpc) galacto-centric distance.

Analogous signatures of intracluster stellar populations have been shown 
previously in the Virgo, Fornax and Coma clusters \citep[see e.g., ][]{Paolillo02, Arnaboldi04, Peng11,
Longobardi15} but the catalog presented here will offer a unique 
chance to perform a fully dynamical analysis of both the bound and unbound GC 
populations \citep{DAbrusco16}.

\section{Conclusions}
\label{sec:conclusions}
In this paper, we presented the results of a wide 1 deg$^2$ spectroscopic survey of 
GCs in the Fornax cluster, focusing on observations, data 
reduction and presentation of the final catalogue. This is the first work of a 
larger multi-instrument program dubbed the Fornax Cluster VLT Spectroscopic 
Survey (FVSS), which, in synergy with the ongoing Fornax Deep Survey (FDS), 
aims to study the formation, evolution and dynamics of galaxies and small 
stellar systems in the Fornax cluster.

Our analysis is based on observations with VST/OmegaCam for imaging and 25 
VST/VIMOS masks for spectroscopic follow-up. Objects were pre-selected 
with multi-band imaging based on optical FDS and near-infrared NGFS photometry, 
supported also by the many spectroscopic catalogues already 
present for this region of the sky.  Approximately 4500 objects were observed. 
The total observation time was 37.5 hours, mostly in sub-arcsec seeing 
conditions. Data reduction was performed using the ESO VIMOS pipeline. 
Redshifts were extracted using {\it iraf/fxcor}, whereas the remaining data analysis 
was carried out using customer {\it python} code. Only the Calcium Triplet region
(8498 to 8662 \AA) was used for redshift estimation, but the quality of the fit 
was assessed using the full spectrum from 4800 to 10000 \AA.

After a visual inspection of candidate spectra, we compiled a consensus 
catalogue of 372 GCs, 15 UCDs (objects with $i \le 20.3$ mag) and 464 Galactic 
stars. Most GCs belong to the dominant galaxy NGC~1399, but we estimated that 
30-40 objects might belong to NGC~1404, NGC~1380, and to other major galaxies 
in the observed field.

We have used the new complete catalog of GCs to derive the total  $v_{\rm rms}$ 
of the GC sample and also split the sample in red and blue subsamples using
the threshold $(g - i) = 0.85$ mag.
We have demonstrated that all profiles show a similar signature at around 
$R\sim10$ arcmin of a kinematical transition from a low $v_{\rm rms}$ regime ($\sim 250 \kms$) to a 
higher one ($\sim 350 \kms$), with the former being consistent with the central velocity 
dispersion of NGC~1399 (the central cluster galaxy) and the latter being 
consistent with the velocity dispersion of the Fornax cluster galaxy population. This 
demonstrates that at $R>10$ arcmin both GC populations feel strongly the cluster 
potential, rather than the increasingly weaker galaxy potential galaxy potential. Parallel to this paper 
we also performed a similar analysis where we present similar evidence based on 
kinematical distribution of planetary nebulae (PNe) in the core of the cluster, 
up to 200 Kpc (Spiniello et al. 2018, submitted). PNe also show a rise in the 
velocity dispersion compatible with PNe tracing the potential of the cluster as 
a whole. A detailed kinematical characterization of these intra cluster populations will 
be the topic of a forthcoming analysis. We point out here that both lines of kinematical 
evidence (GCs and PNe) corroborate the photometric evidence, found in the deep 
galaxy photometry, of a transition region between the bright central galaxy and 
the outer exponential halo at $R\sim10$ arcmin from \citet{Iodice16} and 
show, from the dynamical point of view, the emergence of an intracluster 
population of GCs (as well as PNe) in the Fornax galaxy Cluster.

\section*{Acknowledgments}

NRN and EI acknowledge financial support from the European Union's Horizon
2020 research and innovation programme under the Marie Sk\l{}odowska-Curie
grant agreement No 721463 to the SUNDIAL ITN network and  PRIN INAF 2014 
``Fornax Cluster Imaging and Spectroscopic Deep Survey''.
CT is supported through an NWO-VICI grant (project number 639.043.308).
CSpiniello has received funding from the European Union's Horizon 2020 
research and innovation programme under the Marie Sk\l{}odowska-Curie
actions grant agreement No 664931.
M.P. acknowledges financial contribution from the agreement ASI-INAF No 2017-14-H.O.
T.L. acknowledges financial support from the European Union's Horizon 2020 
research and innovation programme under the Marie Sk\l{}odowska-Curie 
grant agreement No.~721463 to the SUNDIAL ITN network.
P.E. acknowledges support from the Chinese Academy of Sciences (CAS) 
through CAS-CONICYT Postdoctoral  Fellowship CAS150023 administered 
by the CAS South America Center for Astronomy (CASSACA) in Santiago,  Chile.
MS acknowledge finacial support from the VST project (P.I. P. Schipani).
The authors acknowledge financial support from the European Union's 
Horizon 2020 research and innovation programme under Marie Sk\l{}odowska-Curie 
grant agreement No 721463 to the SUNDIAL ITN network.
GvdV acknowledges funding from the European Research Council (ERC) under 
the European Union's Horizon 2020 research and innovation programme 
under grant agreement No 724857 (Consolidator Grant ArcheoDyn).
R.D'A. is supported by NASA contract NAS8-03060 (Chandra X-ray Center).
THP acknowledges support by the FONDECYT Regular Project No. 1161817 and 
the BASAL Center for Astrophysics and Associated Technologies (PFB-06).
This research made use of Astropy, a community-developed core Python 
package for Astronomy (Astropy Collaboration, 2018)

\bibliographystyle{mnras}
\bibliography{ref3}
\end{document}